\RequirePackage{fix-cm}
\documentclass[twocolumn,epjc3,final]{svjour3}  
\smartqed  
 \usepackage{mathrsfs}
\usepackage{amsmath, txfonts}
\usepackage{graphicx,bm,booktabs,bm}
\usepackage{longtable,lscape}
\usepackage{overpic,multirow,color}
\definecolor{cover}{rgb}{0.77,0.87,0.88}
\definecolor{blueone}{rgb}{0.1,0.1,.7}
\definecolor{citec}{rgb}{0.14,0.47,0.09}
\definecolor{two}{rgb}{0.0,0.5,0.}
\definecolor{three}{rgb}{.5,.1,0.15}
\usepackage[bookmarks=true,bookmarksopen=false,plainpages=false,breaklinks=true,
   bookmarksnumbered=true,hypertexnames=false,
   filecolor=blue,urlcolor=three,menucolor=three,
   linkcolor=three,citecolor=blueone, colorlinks,
   anchorcolor=blue,runcolor=pink,frenchlinks=red
   pdfstartview=FitH,pdftitle=title,%
   pdfauthor=author]{hyperref}
              \allowdisplaybreaks[4]

\graphicspath{{Figures/}} %

\journalname{Eur. Phys. J. C}
\begin{document}
\title{Y(4626) as a molecular state from interaction ${D}^*_s\bar{D}_{s1}(2536)-{D}_s\bar{D}_{s1}(2536)$}
\author{Jun He\thanksref{e1,addr1}, Yi Liu\thanksref{addr1}, Jun-Tao Zhu\thanksref{addr1}
\and Dian-Yong Chen\thanksref{addr2}
}                     
\thankstext{e1}{Corresponding author: junhe@njnu.edu.cn}
\institute{Department of  Physics and Institute of Theoretical Physics, Nanjing Normal University,
Nanjing 210097, China\label{addr1}
\and
School of Physics, Southeast University, Nanjing 210094,  China\label{addr2}
}

\date{Received: date / Revised version: date}
%
\maketitle

\abstract{
Recently, a new structure $Y(4626)$ was reported by the Belle Collaboration in the process $e^+e^-\to D_s^+D_{s1}(2536)^-$. In this work, we propose an assignment of the $Y(4626)$ as a ${D}^*_s\bar{D}_{s1}(2536)$ molecular state, which decays into the $D_s^+D_{s1}(2536)^-$ channel through a coupling between ${D}^*_s\bar{D}_{s1}(2536)$ and ${D}_s\bar{D}_{s1}(2536)$ channels.  With the help of the heavy quark symmetry, the potential of the  interaction ${D}^*_s\bar{D}_{s1}(2536)-{D}_s\bar{D}_{s1}(2536)$ is constructed within the one-boson-exchange model, and inserted into the quasipotential Bethe-Salpeter equation.  The pole of obtained scattering amplitude is searched for in the complex-energy plane, which corresponds to a molecular state from the  interaction ${D}^*_s\bar{D}_{s1}(2536)-{D}_s\bar{D}_{s1}(2536)$.  The results suggest that a pole is produced near the ${D}^*_s\bar{D}_{s1}(2536)$ threshold, which exhibits as a peak in the  invariant mass spectrum of the ${D}_s\bar{D}_{s1}(2536)$ channel at about 4626 MeV.  It  obviously favors  the $Y(4626)$ as a ${D}^*_s\bar{D}_{s1}(2536)$  molecular state.  In the same model, other molecular states  from the interaction ${D}^*_s\bar{D}_{s1}(2536)-{D}_s\bar{D}_{s1}(2536)$  are also predicted, which can be checked in future experiments.
} 

\section{Introduction}

Very recently, a new charmoniumlike state $Y(4626)$ was reported  as a structure in the process $e^+e^-\to D_s^+D_{s1}(2536)^-$  based on a sample of 921.9 fb$^{-1}$ accumulated with Belle detector~\cite{Jia:2019gfe}.  The $Y(4626)$ has a mass of $4625.9^{+6.2}_{-6.0}({\rm stat.})\pm0.4({\rm syst})$ MeV and a width of $49.8^{+13.9}_{-11.5}({\rm stat.})\pm 4.0({\rm syst.})$ MeV.  This state is very close to the $Y(4630)$ observed in the process $e^+e^-\to\Lambda_c\bar{\Lambda}_c$~\cite{Pakhlova:2008vn} and also near the $Y(4660)$ observed in the process $e^+e^-\to\gamma_{ISR}\pi^+\pi^-\psi(3686)$~\cite{Wang:2007ea}.  The new observation makes the situation in this energy region more complicated. 

In Ref.~\cite{Tan:2019knr}, the $Y(4626)$ was interpreted as a tetraquark by  a calculation in the constituent quark model. In the other side, there exists many theoretical  explanations of the $Y(4660)$ and $Y(4630)$.  The $Y(4660)$ was suggested to be interpreted as a $5^3S_1$ $c\bar{c}$ state in the conventional  quark model~\cite{Ding:2007rg},  a $f_0\psi'$  bound state~\cite{Guo:2008zg}, or a tetraquark~\cite{Maiani:2014aja,Chen:2010ze}.  Since the  $Y(4630)$ was observed near the  $\Lambda_c\bar{\Lambda}_c$ threshold, it was proposed to be a  $\Lambda_c\bar{\Lambda}_c$ molecular state, which is supported by the strong attraction through $\sigma$ and $\omega$ exchanges calculated in Refs.~\cite{Lee:2011rka,Simonov:2011jc}.  Some authors also suggested that these two states are the same state~\cite{Bugg:2008sk,Cotugno:2009ys,Guo:2010tk}.   If we only consider the mass of the newly observed $Y(4626)$, all the interpretations of the $Y(4660)$ and $Y(4630)$ can be used to explain its origin and internal structure.  To have further understanding about these states, the decay channels should be considered.

Since the $Y(4626)$ was observed in the $D_s^+D_{s1}^-$ channel (here and hereafter, the number 2536 in $D_{s1}(2536)$ will be omitted). It is natural to assume it as a bound state of an anticharm-strange meson and a charm-antistrange meson.  
 Different from the case of $Y(4630)$ which was observed in the $\Lambda_c\bar{\Lambda_c}$ channel and close to the $\Lambda_c\bar{\Lambda_c}$ threshold also,  the $Y(4626)$ is much higher than the  $D_s^+D_{s1}^-$ threshold.   In fact, the ${D}^*_s\bar{D}_{s1}$ threshold is about 4648 MeV, which is a little higher than the mass of  $Y(4626)$. Hence, the $Y(4626)$ can be assigned as a candidate of the ${D}^*_s{D}_{s1}$ molecular state.  In the literature, such molecular states composed of an anticharm-strange meson and a charm-antistrange meson  have been discussed, such as the $Y(4140)$ as a ${D}_s^*\bar{D}_s^*$ state and $Y(4274)$ as a ${D}_s\bar{D}_{s0}$ state~\cite{Liu:2010hf,He:2011ed,Karliner:2016ith}. Besides,  the $Z_c(4430)$ and $Y(4390)$ were also suggested to be states form the ${D}^*\bar{D}_1$ interaction~\cite{He:2017mbh}.  If we consider the ${D}^*\bar{D}_1$ threshold is about 4430 MeV, the mass gap between ${D}^*\bar{D}_1$ and ${D}_s^*\bar{D}_{s1}$ thresholds is about  220 MeV, which is very close to the mass gap between the $Y(4626)$ and the $Y(4390)$.

 Under such assumption, the observation of the $Y(4626)$ in the $D_s^+D_{s1}^-$ channel is also easy to understand. The vector $D^*_s$ meson can be converted into pseudoscalar $D_s$  meson by exchanging an $\eta$ or $\phi$ meson, which leads to the coupling between ${D}^*_s\bar{D}_{s1}$ and ${D}_s\bar{D}_{s1}$ channels.  Hence, in the current work, we will consider  coupled-channel interaction ${D}^*_s\bar{D}_{s1}-{D}_s\bar{D}_{s1}$   in the calculation.  With this interaction, we study the possible bound state from the ${D}^*_s\bar{D}_{s1}$ interaction and its coupling with the ${D}_s\bar{D}_{s1}$ 
channel, where the $Y(4626)$ was observed,  in a quasipotential Bethe-Salpeter equation (qBSE) approach.

This work is organized as follows. After introduction, the  reduction of potential kernel  of coupled-channel interaction ${D}^*_s\bar{D}_{s1}-{D}_s\bar{D}_{s1}$    is presented, which is obtained with the help of the heavy quark symmetry. The relevant  coupling constants are also discussed and given there. And the qBSE approach is introduced briefly. Then, the potential is inserted into the qBSE to search for a pole corresponding  to the $Y(4626)$ and the numerical results will be given in Section 3.  The molecular states with other quantum numbers are also predicted within the same model.  Finally, the article ends with summary  and  discussion.

\section{Theoretical frame}

Since the $Y(4626)$ was observed in the process $e^+e^-\to D_s^+D_{s1}^-$, it should carry quantum numbers $I(J^{PC})=0(1^{--})$. First, we need construct the flavor functions for the  $\bar{D}^*_sD_{s1}-\bar{D}_sD_{s1}$ system with definite $I(J^{PC})$. 
Since only isoscalar state can be formed from the ${D}^*_s\bar{D}_{s1}-{D}_s\bar{D}_{s1}$ system, we need not consider the isospin in construction of  flavor function.  The spin parity $J^P$ will be determined in the partial wave decomposition, which will be explained explicitly later. Here, we give the flavor function for a definite charge parity $C$ as 
\begin{align}
 |D_s\bar{D}_{s1}\rangle&\equiv \frac{1}{\sqrt{2}}\left[|D^+_sD^-_{s1}+CD^-_sD^+_{s1}\rangle\right],\\
 |D^*_s\bar{D}_{s1}\rangle&= \frac{1}{\sqrt{2}}\left[|D^{*+}_sD^-_{s1}-CD^{*-}_sD^+_{s1}\rangle\right].
 \end{align}
Here, we adopt the conventions ${\cal C}D_s^\pm{\cal C}^{-1}=D^\mp_s$, ${\cal C}D_{s1}^\pm{\cal C}^{-1}=D^\mp_{s1}$,  and ${\cal C}D_s^{*\pm}{\cal C}^{-1}=-D^{*\mp}_s$, which are also adopted in the Lagrangians used in the current work.  It is easy to check that the wave functions given above carry a charge parity $C$. 

Besides the flavor function, to study the bound state from the interaction and  coupling between different channels, we need construct the  potential kernel within the one-boson-exchange model, which is widely used to describe the interaction between two hadrons.  Because only charm strange mesons are involved, only $\phi$ and $\eta$ mesons are exchanged in the interaction considered in the current work.  The relevant Lagrangians  will be presented in the following.

\subsection{Relevant Lagrangians}
We need consider the couplings of light mesons to heavy-light anticharmed mesons in $H$ and $T$ doublets.  In terms of heavy quark limit and chiral symmetry,  the Lagrangians has been  constructed  in the literature as \cite{Cheng:1992xi,Yan:1992gz,Wise:1992hn,Casalbuoni:1996pg,Ding:2008gr},
\begin{eqnarray}
  \label{eq:lag}
  \mathcal{L} &=&
ig\langle H_b{\cal
A}\!\!\!\slash_{ba}\gamma_5\overline{H}_a+ig\langle\overline{\tilde{H}}_a{\cal
A}\!\!\!\slash_{ab}\gamma_5\tilde{H}_b\rangle\rangle\nonumber\\
&+& i\beta\langle H_bv^{\mu}({\cal
V}_{\mu}-V_{\mu})_{ba}\overline{H}_a\rangle+i\lambda\langle
H_{b}\sigma^{\mu\nu}F_{\mu\nu}(V)_{ba}\overline{H}_a\rangle
\nonumber\\
&-&i\beta\langle
\overline{\tilde{H}}_av^{\mu}({\cal
V}_{\mu}-V_{\mu})_{ab}\tilde{H}_b\rangle+i\lambda\langle
\overline{\tilde{H}}_a\sigma^{\mu\nu}F_{\mu\nu}(V)_{ab}\tilde{H}_b\rangle\nonumber\\
&+&ik\langle T^{\mu}_{b}{\cal A}\!\!\!\slash_{ba}\gamma_5\overline{T}_{a\mu}\rangle+ik\langle\overline{\tilde{T}}^{\mu}_a{\cal
A}\!\!\!\slash_{ab}\gamma_5\tilde{T}_{b\mu}\rangle\nonumber\\
&+& i\beta_2\langle
T^{\lambda}_bv^{\mu}({\cal
V}_{\mu}-V_{\mu})_{ba}\overline{T}_{a\lambda}\rangle+i\lambda_2\langle
T^{\lambda}_b\sigma^{\mu\nu}F_{\mu\nu}(V)_{ba}\overline{T}_{a\lambda}\rangle\nonumber\\
&-&i\beta_2\langle\overline{\tilde{T}}_{a\lambda}v^{\mu}({\cal V}_{\mu}-V_{\mu})_{ab}\tilde{T}^{\lambda}_{b}\rangle+i\lambda_2\langle\overline{\tilde{T}}_{a\lambda}\sigma^{\mu\nu}F_{\mu\nu}(V)_{ab}\tilde{T}^{\lambda}_{b}\rangle\nonumber\\
&+&i\frac{h_1}{\Lambda_{\chi}}\langle
T^{\mu}_b(D_{\mu}{\cal
A}\!\!\!\slash)_{ba}\gamma_5\overline{H}_a\rangle+i\frac{h_2}{\Lambda_{\chi}}\langle
T^{\mu}_b(D\!\!\!\!/ {\cal
A}_{\mu})_{ba}\gamma_5\overline{H}_a\rangle\nonumber\\
&+&i
\frac{h_1}{\Lambda_{\chi}}\langle\overline{\tilde{H}}_a({\cal
A}\!\!\!\slash\stackrel{\leftarrow}{D_{\mu}'})_{ab}\gamma_5\tilde{T}^{\mu}_b\rangle+i\frac{h_2}{\Lambda_{\chi}}\langle\overline{\tilde{H}}_a({\cal
A}_{\mu}\stackrel{\leftarrow}{D\!\!\!\slash'})_{ab}\gamma_5\tilde{T}^{\mu}_b\rangle\nonumber\\
&+&i\zeta_1\langle
T^{\mu}_b({\cal V}_{\mu}-V_{\mu})_{ba}\overline{H}_a\rangle+\mu_{1}\langle
T^{\mu}_b\gamma^{\nu}F_{\mu\nu}(V)_{ba}\overline{H}_a\rangle\nonumber\\
&-&i\zeta_1\langle\overline{\tilde{H}}_a({\cal
V}_{\mu}-V_{\mu})_{ab}\tilde{T}^{\mu}_b\rangle+\mu_1\langle\overline{\tilde{H}}_a\gamma^{\nu}F_{\mu\nu}(V)_{ab}\tilde{T}^{\mu}_b\rangle\nonumber\\ &+& {\rm H.\ c.}.\label{L0}
\end{eqnarray}
 where $v=(1,\mathbf{0})$, and
the axial current is
${\cal A}^\mu=\frac{1}{2}(\xi^\dag\partial_\mu\xi-\xi \partial_\mu
\xi^\dag)=\frac{i}{f_\pi}\partial_\mu{\mathbb P}+\cdots$ with
$\xi=\exp(i\mathbb{P}/f_\pi)$ and $f_\pi=132$
MeV.
Vector current $\mathcal{V}_\mu=\frac{i}{2}[\xi^\dag(\partial_\mu\xi)
+(\partial^\mu\xi)\xi^\dag]=0$ with $V^\mu_{ba}=ig_V\mathbb{V}^\mu_{ba}/\sqrt{2}$, and
$F_{\mu\nu}(V)=\partial_\mu V_\nu - \partial_\nu V_\mu +
[V_\mu,{\ } V_\nu]$. The $\mathbb
P$ and $\mathbb V$ are the pseudoscalar and vector matrices as
\begin{eqnarray}
    {\mathbb P}&=&\left(\begin{array}{ccc}
        \frac{1}{\sqrt{2}}\pi^0+\frac{\eta}{\sqrt{6}}&\pi^+&K^+\\
        \pi^-&-\frac{1}{\sqrt{2}}\pi^0+\frac{\eta}{\sqrt{6}}&K^0\\
        K^-&\bar{K}^0&-\frac{2\eta}{\sqrt{6}}
\end{array}\right),\nonumber\\
\mathbb{V}&=&\left(\begin{array}{ccc}
\frac{\rho^{0}}{\sqrt{2}}+\frac{\omega}{\sqrt{2}}&\rho^{+}&K^{*+}\\
\rho^{-}&-\frac{\rho^{0}}{\sqrt{2}}+\frac{\omega}{\sqrt{2}}&K^{*0}\\
K^{*-}&\bar{K}^{*0}&\phi
\end{array}\right),\label{MPV}
\end{eqnarray}
which correspond to $(D^0,D^+,D_s^+)$ 
and  $(\bar{D}^0,D^-,D_s^-)$.

To constrain the interaction, the values of  coupling constants involved should be determined. The coupling constants for the $H$ doublet are relatively well determined in the literature with the heavy quark symmetry and available experimental data,  i. e., $g=0.59$, $\beta=0.9$, $\lambda=0.56$~GeV$^{-1}$ with $g_V=5.9$ and $f_\pi=0.132$ GeV~\cite{Chen:2019asm,Liu:2011xc,Isola:2003fh,Falk:1992cx}.  For the couplings with the $T$ doublet involved, some coupling constants were also determined in the literature.   Casalbuoni and coworkers extracted  $(h_1+h_2)/\Lambda_\chi=0.55$~GeV$^{-1}$ for experimental information~\cite{Casalbuoni:1996pg}. Falk and Luke obtained an approximate relation $k=g$ by a quark model calculation~\cite{Falk:1992cx}.  In Ref.~\cite{Chen:2019asm}, the $k$ are related to the coupling constant for the $\pi NN$ vertex by comparing the results in hadronic and quark levels, and a relation was reached as $k/f_\pi=3\sqrt{2}g_{\pi NN}/(10 m_N)$ with $g^2_{\pi NN}/4\pi=13.60$, which leads to $k=0.78$.  Such value is close to the approximation $k=g$.  Here, we still use $k=g=0.59$ as adopted in Ref.~\cite{Chen:2019asm}.   

Analogously,  the values of $\beta_2$ and $\lambda_2$ were determined also in Ref.~\cite{Chen:2019asm} as  $\beta_2g_V=-2g_{\rho NN}$ with $g^2_{\rho NN}/4\pi=0.84$, which leads to $\beta_2=1.10$, and $\lambda_2 g_V=3(g_{\rho NN}+f_{\rho NN})/(10 m_N)$ with $k_\rho=6.10$, which leads to $\lambda_2=-1.25$~GeV$^{-1}$.  The $\beta_2$ obtained there is close to the value in Ref.~\cite{Dong:2019ofp} where $g_{D_1D_1V}\approx g_{DDV}$ was adopted, which leads to $\beta_2=\beta=0.9$.  For the $\lambda_2$, the value in Ref.~\cite{Chen:2019asm} is dependent on the coupling constants $g_{\rho NN}$ and $k_\rho$. The value of $g_{\rho NN}$ is usually consistent to each other in the literature while the $k_\rho$ has two suggested values, about 6 and about 1, respectively~\cite{Matsuyama:2006rp,Penner:2002md,Machleidt:1987hj}.  In Ref.~\cite{Dong:2019ofp}, such term was omitted, which corresponds to $\lambda_2=0$~GeV$^{-1}$. In the current work, we will choose $\beta_2=1.1$ and $\lambda_2=-0.6$~GeV$^{-1}$. 
The coupling constants $\mu_1$ and $\zeta_1$ are not well determined.   In Ref.~\cite{Dong:2019ofp}, the authors made an approximation as $\mu_1=0$~GeV$^{-1}$ and $\zeta_1=-0.04\sim-0.25$ from the decay widths of the $K_1(1270)$ and the $K_1(1400)$ into $\rho N$ channel. In the current work, we adopt $\mu_1=0$~GeV$^{-1}$ as in Ref.~\cite{Dong:2019ofp}. The values of the $\lambda_2$ and $\zeta_1$ will be discussed explicitly later. 

\normalsize
The $H$ and $T$ doublet fields are defined as 
\begin{eqnarray}
  \label{eq:doublets}
  H_a &=& \frac{1+\rlap\slash{v}}{2}[P_{a\mu}^{*}\gamma^\mu -P_a\gamma_5],\quad
  \bar{H}_b =
  [P^{*\dag}_{a\mu}\gamma^\mu +P_a^\dag\gamma_5]\frac{1+\slash \!\!\! v}{2},\nonumber\\
  \tilde{H}_a &=& [\tilde{P}_{a\mu}^{*}\gamma^\mu-\tilde{P}_a\gamma_5]\frac{1-\rlap\slash{v}}{2},\quad
  \bar{\tilde{H}}_b = \frac{1-\slash
  \!\!\! v}{2}[\tilde{P}^{*\dag}_{a\mu}\gamma^\mu + \tilde{P}_a^\dag\gamma_5],\nonumber\\
  T^{\mu}_a&=&\frac{1+\not
v}{2}[P^{*\mu\nu}_{2a}\gamma_{\nu}-\sqrt{\frac{3}{2}}P_{1a\nu}\gamma_5(g^{\mu\nu}-\frac{1}{3}\gamma^{\nu}(\gamma^{\mu}-v^{\mu}))]
  ,\nonumber\\
    \bar{T}^{\mu}_a&=&[P^{*\mu\nu\dag}_{2a}\gamma_{\nu}+\sqrt{\frac{3}{2}}P^{\dag}_{1a\nu}(g^{\mu\nu}-\frac{1}{3}(\gamma^{\mu}-v^{\mu})\gamma^{\nu})\gamma_5]\frac{1+\not
v}{2}
  ,\nonumber\\
\tilde{T}^{\mu}_{a}&=&[\tilde{P}^{\mu\nu}_{2a}\gamma_{\nu}-\sqrt{\frac{3}{2}}\tilde{P}_{1a\nu}\gamma_5(g^{\mu\nu}-\frac{1}{3}(\gamma^{\mu}-v^{\mu})\gamma^{\nu})]\frac{1-\not
v}{2}  ,\nonumber\\
\bar{\tilde{T}}^{\mu}_{a}&=&\frac{1-\not
v}{2}[\tilde{P}^{\mu\nu\dag}_{2a}\gamma_{\nu}+\sqrt{\frac{3}{2}}\tilde{P}^\dag_{1a\nu}(g^{\mu\nu}-\frac{1}{3}\gamma^{\nu}(\gamma^{\mu}-v^{\mu}))\gamma_5],\label{HT}
\end{eqnarray}
where the $P$ and $P^*$ satisfy the normalization relations $\langle
0|P|{Q}\bar{q}(0^-)\rangle=\sqrt{M_P}$ and $\langle
0|P^*_\mu|{Q}\bar{q}(1^-)\rangle=\epsilon_\mu\sqrt{M_{P^*}}$. Other mesons have analogous normalization relations.

After expanding Eqs~(\ref{L0}) and (\ref{HT}), the effective Lagrangians read,
\begin{eqnarray}\label{eq:lag-p-exch}
  \mathcal{L}_{P^*P^*\mathbb{P}}   &=&
  \frac{2g}{f_\pi}\epsilon_{\mu\nu\alpha\beta} \left(P^{*\mu}_b P^{*\nu\dag}_a+\tilde{P}^{*\mu}_a\tilde{P}^{*\nu\dag}_b\right)v^\alpha \partial^\beta\mathbb{P}_{ba},\nonumber\\
  \mathcal{L}_{P^*P^*\mathbb{V}} &=& \sqrt{2}\beta g_V
  \left( P^*_b\cdot P_a^{*\dag} -\tilde{P}^*_a\cdot\tilde{P}_b^{*\dag}\right)~v\cdot\mathbb{V}_{ba}\nonumber\\
&-&i2\sqrt{2}\lambda
  g_V \left(P^{*\mu}_b P^{*\nu\dag}_a-\tilde{P}^{*\mu}_a\tilde{P}^{*\nu\dag}_b\right)(\partial_\mu\mathbb{V}_\nu-\partial_\nu\mathbb{V}_\mu)_{ba},
  \nonumber\\
    \mathcal{L}_{P_1P_1\mathbb{P}}
  &=&-\frac{5k}{3f_\pi}i\epsilon_{\mu\nu\alpha\beta}\left(P^{\mu}_{1b}
  P^{\nu\dag}_{1a}+ \tilde{P}^{\mu}_{1a}\tilde{P}^{\nu\dag}_{1b}\right)v^\alpha\partial^\beta \mathbb{P}_{ba},\nonumber\\
  \mathcal{L}_{P_1P_1\mathbb{V}}
  &=& \sqrt{2}\beta_2 g_V \left({P}_{1b}\cdot{P}^{\dag}_{1a}-\tilde{P}_{1a}\cdot \tilde{P}^{\dag}_{1b}\right)~v\cdot \mathbb{V}_{ba}\nonumber\\
  &+&\frac{5\sqrt{2}i\lambda_2 g_V}{3}\left({P}^\mu_{1b}{P}^{\nu\dag}_{1a}-\tilde{P}^\mu_{1a}\tilde{P}^{\nu\dag}_{1b}\right)(\partial _\mu\mathbb{V}_{\nu}-\partial_\nu\mathbb{V}_\mu)_{ba},\nonumber\\
    \mathcal{L}_{P^*P\mathbb{P}} &=&
  -\frac{2g}{f_\pi} \left(P_bP^{*\dag}_{a\lambda}+P^*_{b\lambda}P^\dag_{a}-\tilde{P}_a\tilde{P}^{*\dag}_{b\lambda}-\tilde{P}^*_{a\lambda}\tilde{P}^\dag_{b}\right)\partial^\lambda\mathbb{P}_{ba},\nonumber\\
    \mathcal{L}_{P^*P\mathbb{V}} &=&
  -2\sqrt{2}\lambda g_V\varepsilon_{\lambda\alpha\beta\mu}v^\lambda
  \partial^\alpha\mathbb{V}^\beta_{ba}\nonumber\\
 &\cdot& \left(P_bP^{*\mu\dag}_a 
  +P_b^{*\mu}P^\dag_a + \tilde{P}_a\tilde{P}^{*\mu\dag}_b
   +\tilde{P}_a^{*\mu}\tilde{P}^\dag_b
 \right),\nonumber\\
	\mathcal{L}_{PP\mathbb{V}} &=& -\sqrt{2}\beta g_V\left(P_b P_a^\dag -\tilde{P}_a\tilde{P}_b^\dag \right)~v\cdot\mathbb{V}_{ba}, \nonumber\\
  \mathcal{L}_{P_1P\mathbb{P}}
  &=&0,\nonumber\\
    \mathcal{L}_{P_1P\mathbb{V}}
  &=&\frac{2g_V}{\sqrt{3}}\left[\zeta_1(P_bP_{1a}^\dag +P_{1b}P^\dag_a-\tilde{P}_a\tilde{P}_{1b}^\dag -\tilde{P}_{1a}\tilde{P}^\dag_b)\cdot\mathbb{V}_{ba}\right.\nonumber\\
  & +&\left.i\mu_1( P_bP_{1a}^{\mu\dag} +P_{1b}^\mu P^\dag_a-\tilde{P}_a\tilde{P}_{1b}^{\mu\dag} -\tilde{P}_{1a}^\mu \tilde{P}^\dag_b)\right.\nonumber\\
 &\cdot& \left.v^\nu(\partial_\mu\mathbb{V}_\nu-\partial_\nu\mathbb{V}_\mu)_{ba} \right],\nonumber\\
  \mathcal{L}_{P_1P^*\mathbb{P}}
  &=& -\sqrt{\frac{2}{3}}\frac{h_1+h_2}{f_\pi\Lambda}\nonumber\\
  &\cdot&\left[3(P^\mu_{1b}P^{*\dag}_{\nu a}+P^{*}_{\nu b}P^{\mu\dag}_{1a}+\tilde{P}^\mu_{1a}\tilde{P}^{*\dag}_{\nu b}+\tilde{P}^{*}_{\nu a}\tilde{P}^{\mu\dag}_{1b})\partial^\mu\partial_\nu \mathbb{P}_{ba}\right.\nonumber\\
  &-&(P_{1b}\cdot P^{*\dag}_{ a}+P^{*}_{ b}\cdot P^{\dag}_{1a}+\tilde{P}_{1a}\cdot \tilde{P}^{*\dag}_{ b}+\tilde{P}^{*}_{ a}\cdot \tilde{P}^{\dag}_{1b})\partial^2\mathbb{P}_{ba}\nonumber\\
  &+&\frac{1}{m_{P_1}m_{P^*}}(\partial^\nu P^\mu_{1b}\partial^\lambda P^{*\dag}_{a\mu}+\partial^\nu P^{*\mu}_{b}\partial^\lambda P^{\dag}_{1a\mu}\nonumber\\
  &+&\left.\partial^\nu \tilde{P}^\mu_{1a}\partial^\lambda \tilde{P}^{*\dag}_{b\mu}+\partial^\nu \tilde{P}^{*\mu}_{a}\partial^\lambda \tilde{P}^{\dag}_{1b\mu})\partial_\nu\partial_\lambda \mathbb{P}_{ba}\right],
  \nonumber\\
  \mathcal{L}_{P_1P^*\mathbb{V}}
  &=& \frac{g_V}{\sqrt{3}}\epsilon_{\mu\nu\alpha\beta}(P^\mu_{1b}P^{*\nu\dag}_a +P^{*\mu}_b P_{1a}^{\nu\dag}+\tilde{P}^\mu_{1a}\tilde{P}^{*\nu\dag}_b +\tilde{P}^{*\mu}_a \tilde{P}_{1b}^{\nu\dag}) \nonumber\\
  &\cdot&(2\mu_1\partial^\alpha+i\zeta_1 v^\alpha)\mathbb{V}^\beta_{ba},\label{L}
\end{eqnarray}
where  the $v$ should be replaced by $i\overleftrightarrow{\partial}/\sqrt{m_im_f}$ with the $m_{i,f}$ being the mass of the initial or final heavy mesons.

\subsection{ The potential kernel and the qBSE approach}

\begin{figure}[h!]
\centering
\includegraphics[bb=70 680 600 770, clip,scale=0.52]{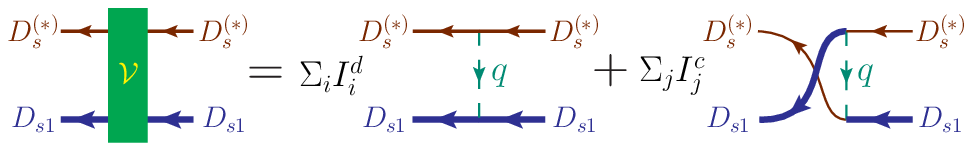}
\caption{The diagrams for the direct (left) and cross (right) potentials. The thin (brown) and thick (blue) lines are for $D^{(*)}$ and $D_{s1}$ mesons, respectively.}
\label{V}
\end{figure}

The potential of the interaction ${D}^*_s\bar{D}_{s1}-{D}_s\bar{D}_{s1}$   is constructed with the help of the vertices which can be easily obtained from   above Lagrangians.  As discussed in Refs.~\cite{He:2014nya,He:2015mja}, there are two types of diagrams, which include direct and cross ones as illustrated in Fig.~\ref{V}.  

As in our previous work~\cite{He:2019rva}, we do not give the explicit form of the potential in this work. Instead, we input  the vertices $\Gamma$  into the code directly and the potential can be obtained as
\begin{eqnarray}%
{\cal V}_{\eta}=I_\eta\Gamma_1\Gamma_2 P_{\eta}(q^2),\quad
{\cal V}_{\phi}=I_\phi\Gamma_{1\mu}\Gamma_{2\nu}  P^{\mu\nu}_{\phi}(q^2),
\label{Eq:V}
\end{eqnarray}
where  the propagators of the exchanged light mesons are also needed, which read,
\begin{eqnarray}%
P_\eta(q^2)&=& \frac{i}{q^2-m_\eta^2},\nonumber\\
P^{\mu\nu}_\phi(q^2)&=&i\frac{-g^{\mu\nu}+q^\mu q^\nu/m^2_{\phi}}{q^2-m_\phi^2},
\end{eqnarray}
where the $q$ is the momentum of the exchanged meson and $m_\eta$ and $m_\phi$ are the masses of the $\eta$ and $\phi$ mesons.
The  $I_{(\eta,\phi)}$  is the flavor factor for $\eta$ or $\phi$ meson exchange, which may be different for direct and cross diagrams. It can be derived with the Lagrangians in Eq.~(\ref{L}) and the matrices in Eq.~(\ref{MPV}). The explicit values are listed in Table~\ref{flavor factor}.

\renewcommand\tabcolsep{0.39cm}
\renewcommand{\arraystretch}{2}
\begin{table}[h!]
\caption{The flavor factors for certain meson exchanges of certain interaction. The $C$ is the charge parity.  \label{flavor factor}}
\begin{tabular}{c|rrrr}\bottomrule[2pt]
& $I^d_\eta $&$I^c_\eta $ & $I^d_\phi $ & $I^c_\phi $ \\\hline
${D}_s\bar{D}_{s1}\to {D}_s\bar{D}_{s1}$&$2/3$ &$2/3C$ &$1$&$C$\\
 ${D}^*_s\bar{D}_{s1}\to{D}^*_s \bar{D}_{s1}$&$2/3$ &$-2/3C$ &$1$&$-C$\\
${D}^*_s\bar{D}_{s1}\to {D}_s\bar{D}_{s1}$&$-2/3$ &$-2/3C$ &$-1$&$-C$\\
${D}_s\bar{D}_{s1}\to {D}^*_s\bar{D}_{s1}$&$-2/3$ &$2/3C$ &$-1$&$C$\\
\toprule[2pt]
\end{tabular}

\end{table}

To calculate the scattering amplitude, the obtained potential  can be inserted into the qBSE, which was decomposed on spin parity $J^P$ as~\cite{He:2018plt,He:2016pfa,He:2017aps,He:2015yva,He:2015cea},
\begin{eqnarray}
i{\cal M}^{J^P}_{\lambda'\lambda}({\rm p}',{\rm p})
&=&i{\cal V}^{J^P}_{\lambda',\lambda}({\rm p}',{\rm
p})+\sum_{\lambda''}\int\frac{{\rm
p}''^2d{\rm p}''}{(2\pi)^3}\nonumber\\
&\cdot&
i{\cal V}^{J^P}_{\lambda'\lambda''}({\rm p}',{\rm p}'')
G_0({\rm p}'')i{\cal M}^{J^P}_{\lambda''\lambda}({\rm p}'',{\rm
p}),\label{Eq: BS_PWA}
\end{eqnarray}
where the sum extends only over nonnegative helicity $\lambda''$ because only the independent helicity amplitudes are considered in the calculation.  Here we adopt the covariant spectator approximation to reduce the Bethe-Salpeter equation in to a qBSE, which leads to a reduced propagator in the center-of-mass frame with $P=(W,{\bm 0})$ as~\cite{He:2011ed,He:2015mja,He:2013oma,Gross:2008ps}
\begin{align}
	G_0&=\frac{\delta^+(p''^{~2}_h-m_h^{2})}{p''^{~2}_l-m_l^{2}}
	\nonumber\\&=\frac{\delta^+(p''^{0}_h-E_h({\rm p}''))}{2E_h({\rm p''})[(W-E_h({\rm
p}''))^2-E_l^{2}({\rm p}'')]}.
\end{align}
As required by the spectator approximation, the heavier particle (remarked with $h$, $D_{s1}$ here)  is on shell, which satisfies  $p''^0_h=E_{h}({\rm p}'')=\sqrt{
m_{h}^{~2}+\rm p''^2}$. The $p''^0_l$ for the lighter particle (remarked as $l$, $D_s$ and $D^*_s$ here) is then $W-E_{h}({\rm p}'')$. Here and hereafter, a definition ${\rm p}=|{\bm p}|$ will be adopted. 

The dynamical mechanism of our model is introduced in the potential kernel ${\cal V}$. After The partial wave decomposition,  the potential obtained in Eq.~(\ref{Eq:V}) can be related to the  ${\cal V}^{J^P}$ with fixed spin parity used in Eq.~(\ref{Eq: BS_PWA}) as
\begin{eqnarray}
{\cal V}_{\lambda'\lambda}^{J^P}({\rm p}',{\rm p})
&=&2\pi\int d\cos\theta
~[d^{J}_{\lambda\lambda'}(\theta)
{\cal V}_{\lambda'\lambda}({\bm p}',{\bm p})\nonumber\\
&+&\eta d^{J}_{-\lambda\lambda'}(\theta)
{\cal V}_{\lambda'-\lambda}({\bm p}',{\bm p})],
\end{eqnarray}
where the factor $\eta=PP_lP_h(-1)^{J-J_l-J_h}$ with $P$ and $J$ being parity and spin for system, $D^{(*)}_s$ meson or $D_{s1}$ meson. The initial and final relative momenta are chosen as ${\bm p}=(0,0,{\rm p})$  and ${\bm p}'=({\rm p}'\sin\theta,0,{\rm p}'\cos\theta)$. The $d^J_{\lambda\lambda'}(\theta)$ is the Wigner d-matrix. 

Now we need treat an integral equation,  to avoid divergence,  form factor for the off-shell particle is usually introduced. In the qBSE approach,  we usually adopt an  exponential form factor into the  propagator as
\begin{eqnarray}
G_0({\rm p})\to G_0({\rm p})\left[e^{-(k_l^2-m_l^2)^2/\Lambda^4}\right]^2,\label{regularization}
\end{eqnarray} 
where $k_l$ and $m_l$ are the momentum and mass of  the lighter one of two heavy mesons.  For the exchanged meson, we also introduce a exponential form factor as  $F(q^2)=e^{-(m_e^2-q^2)^2/\Lambda^2}$
with $m_e$ and $q$ being the mass and momentum of the exchanged light meson. Here the cutoffs in all form factors are chosen as the same for simplification. We would like to note that in our approach we keep covariant from factors  without nonrelativistc approximation as done in the propagator, which is the characteristic of  covariant-spectator quasipotential approximation adopted in the current work~\cite{Gross:2008ps}.

To solve the integral equation, we discrete the momenta ${\rm p}$,
${\rm p}'$ and ${\rm p}''$ by the Gauss quadrature with wight $w({\rm
p}_i)$ and have~\cite{He:2011ed,He:2013oma},
\begin{eqnarray}
{M}_{ik}
&=&{V}_{ik}+\sum_{j=0}^N { V}_{ij}G_j{M}_{jk}.
\end{eqnarray}
The above equation is obviously a matrix equation. The index for the helicity can also be included to do the calculation. The matrix element for $j=0$  corresponds to on-shell case. The discreted propagator is written as 
\begin{eqnarray}
	G_{j>0}&=&\frac{w({\rm p}''_j){\rm p}''^2_j}{(2\pi)^3}G_0({\rm
	p}''_j), \nonumber\\
G_{j=0}&=&-\frac{i{\rm p}''_o}{32\pi^2 W}+\sum_j
\left[\frac{w({\rm p}_j)}{(2\pi)^3}\frac{ {\rm p}''^2_o}
{2W{({\rm p}''^2_j-{\rm p}''^2_o)}}\right],
\end{eqnarray}
where the $p_o$ is the on-shell momentum in the center of mass frame. 

In the current work, we will present the effect of the $1^-$ bound state from the interaction on the invariant mass spectrum of the ${D}_s\bar{D}_{s1}$ channel.  Since we do not consider the initial $e^+e^-$ collision explicitly, the invariant mass
distribution is given approximately as 
\begin{eqnarray}
	{d\sigma}/{dW}=C{\rm p}_f|M_{\bar{D}_sD_{s1}\to\bar{D}_sD_{s1}}|^2, \label{Eq: mass}
\end{eqnarray} 
where $C$ is a scale constant and ${\rm p}_f$ is momentum of the final state in the center of mass frame.
The initial and final particles should be on-shell. 
The scattering amplitude is
\begin{eqnarray}
	M_{\bar{D}_sD_{s1}\to\bar{D}_sD_{s1}}=M_{00}=\sum[(1-{ V} G)^{-1}]_{0j}V_{j0}.
\end{eqnarray}
The pole can be searched by variation of $z$ to satisfy $|1-V(z)G(z)|=0$
where  $z=E_R-i\Gamma/2$ being the meson-baryon energy $W$ at the real axis.

\section{The states from the interaction ${D}^*_s\bar{D}_{s1}-{D}_s\bar{D}_{s1}$}

\begin{figure}[h!]
\centering
\includegraphics[scale=1.12]{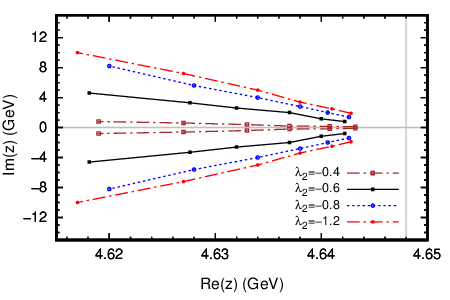}
\caption{The trajectory of pole of bound state  from the ${D}^*_s\bar{D}_{s1}-{D}_s\bar{D}_{s1}$ interaction with $J^{PC}=1^{--}$ in the complex-energy plane of $z=E_R-i\Gamma/2$ with variation of the cutoff $\Lambda$, 3.2 to 3.45 GeV with a step 0.05 GeV from right to left, and different values of the parameter $\lambda_2$ in  unit of GeV$^{-1}$.  \label{para}}
\end{figure}

With above preparation, now, we can scan the scattering amplitude in the complex-energy plane to search for the pole which corresponds to a molecular state.  First, we check the effect of the parameters on our results. As discussed in above section, the coupling constants  $\lambda_2$ and $\zeta_1$ are not well determined in the literature, and the cutoff $\Lambda$ is the free parameter in our model.  With a numerical calculation, it is found that the results are not sensitive to the $\zeta_1$. Hence, in the following, we fix the parameter $\zeta_1$ at a value of $-0.1$.  Now we need to consider the different values of $\lambda_2$, which are in a range from 0 to $-1.2$ GeV$^{-1}$ as discussed in the above section. In Fig.~\ref{para}, we present the moving of the pole for quantum number $J^{PC}=1^{--}$, which can be related to the $Y(4626)$ on which we focus in this work,  in the complex-energy plane with variation of the cutoff $\Lambda$ and different values of $\lambda_2$.

With all values of $\lambda_2$ considered here, the pole is  produced from the interaction ${D}^*_s\bar{D}_{s1}-{D}_s\bar{D}_{s1}$  at a cutoff about 3.2 GeV. The variation  of the value of the $\lambda_2$ effects a little on the real part of the position of the pole, which corresponds to the mass of the molecular state.  More obvious changes can be seen in the imaginary part of the pole, which corresponds to the decay width of the molecular state.  The $\lambda_2=-0.4$~GeV$^{-1}$ leads to a very small width, smaller than 1 MeV with cutoff $\Lambda$ from 3.25 to 3.45~GeV. With the increase of the $|\lambda_2|$, the pole moves further from the real axis, which indicates larger width. The results also suggested that the pole moves to real axis with the decrease of the cutoff $\Lambda$. More calculations suggest that the pole will meet the real axis with $\lambda_2$ about -0.1~GeV$^{-1}$ and leave the real axis again with continuous decrease (we do not give the results with such small $\lambda_2$ in figure to avoid mixing of the curves.). Such results suggest that the second term in the Lagrangian for the $P_1P_1\phi$ vertex  effects the ${D}^*_s\bar{D}_{s1}$ interaction small while has larger effect on the coupling between the ${D}^*_s\bar{D}_{s1}$ and ${D}_s\bar{D}_{s1}$ channels especially with a small $\lambda_2$. If we choose a larger $|\lambda_2|$,  the different choices of the $\lambda_2$ give  qualitatively similar  results. 
In the following, we choose a value of $\lambda_2=-0.6$~GeV$^{-1}$, which corresponds to a larger $k_\rho$ for the $\rho NN$ coupling. 

With the increase of the cutoff $\Lambda$, the pole moves further from both the threshold and the real axis.  It reflects that both ${D}^*_s\bar{D}_{s1}$ interaction  and coupling between two channels are enhanced with a larger cutoff.  The observed mass of the $Y(4626)$ can be reproduced at cutoff $\Lambda=3.4$ GeV, which favors that the $Y(4626)$ state can be related to a ${D}^*_s\bar{D}_{s1}$ molecular state with $1^{--}$. However, the width obtained from the current two-channel calculation is considerably smaller than the experimentally suggested value at Belle.  Even with a larger $|\lambda_2|$ of 1.5, the width $\Gamma=-2{\rm Im}z=20$ MeV, which is still smaller than the experimental value, about 50 MeV~\cite{Jia:2019gfe}. It suggests that  other decay channels including the three-body channels maybe provide considerable width to the $Y(4626)$.

\begin{figure}[h!]
\centering
\includegraphics[scale=0.75]{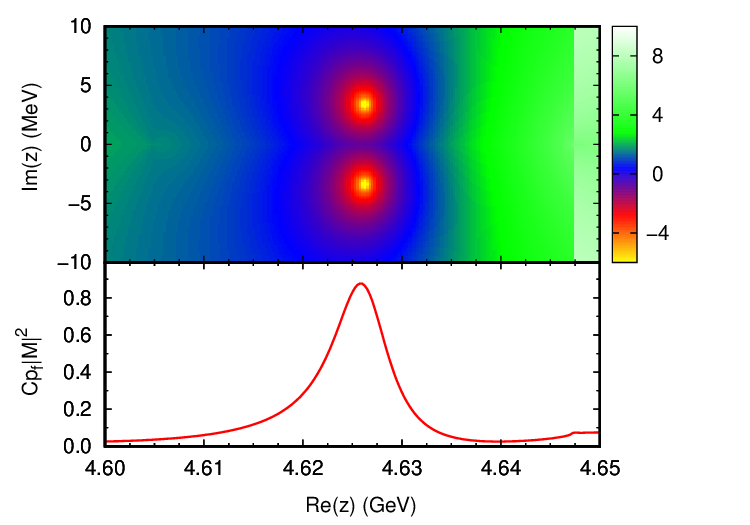}
\caption{The pole of bound state  (upper) and the invariant mass spectrum (lower) from the interaction ${D}^*_s\bar{D}_{s1}-{D}_s\bar{D}_{s1}$  with $J^{PC}=1^{--}$ at $\Lambda=3.4$ GeV. The  color  means the value of $\log|1-V(z)G(z)|$ as shown in the color box.\label{pole}}
\end{figure}

To give a more clearly image of the results, we present the explicit results for the pole from the interaction ${D}^*_s\bar{D}_{s1}-{D}_s\bar{D}_{s1}$  at $\Lambda=3.4$ GeV in Fig.~\ref{pole}.  The pole can be found at $z=4626-3.4i$ MeV which is very close to the experimental mass of the $Y(4626)$.  The peak corresponding to this state can be seen obviously in the ${D}_s\bar{D}_{s1}$ channel.  

The $1^{--}$ state from the interaction ${D}^*_s\bar{D}_{s1}-{D}_s\bar{D}_{s1}$  at 4626 MeV can be related to the $Y(4626)$ observed at Belle. In the same theoretical frame, we could predict other possible molecular state with other quantum numbers. Besides, the pole may be also found near the lower threshold for the ${D}_s\bar{D}_{s1}$ channel. Hence, here we consider all systems from the interaction ${D}^*_s\bar{D}_{s1}-{D}_s\bar{D}_{s1}$ with the quantum numbers which are allowed in S wave, that is,  ${D}^*_s\bar{D}_{s1}(0^{-\pm})$, ${D}^*_s\bar{D}_{s1}(1^{-\pm})$, ${D}^*_s\bar{D}_{s1}(2^{-\pm})$, and ${D}_s\bar{D}_{s1}(1^{-\pm})$.  First, we make a prediction with a  strict condition.  We only consider the bound states with cutoff in a range from 3.0 to 3.55 GeV which is a little larger than that in Fig.~\ref{para}. The results are listed in the following Table~\ref{Tab: bound state1}. Here we only consider the states with a binding energy smaller than 50 MeV because the molecular state is a loosely bound state.

\renewcommand\tabcolsep{0.82cm}
\renewcommand{\arraystretch}{1.6}
\begin{table}[h!]
\begin{center}
\caption{The bound states from the interaction ${D}^*_s\bar{D}_{s1}-{D}_s\bar{D}_{s1}$ at different cutoffs $\Lambda$  in a range from 3.0 to 3.55 GeV.  Here, ``$--$" means that no bound state is found or the bound state has a binding energy larger than 50 MeV.
The cutoff $\Lambda$  and  position of the pole $z$ are in units of GeV and MeV, respectively. \label{Tab: bound state1}
\label{diagrams}}
	\begin{tabular}{c|cc}\bottomrule[1.5pt]
$\Lambda$	&${D}^*_s\bar{D}_{s1}(1^{--})$&${D}^*_s\bar{D}_{s1}(1^{-+})$\\\hline
 3.00 &  $--$&  $--$\\
 3.05 &  $4646-0.1i$&  $--$\\
 3.10   & $4645-0.4i$       & $4648-1.8i$     \\
 3.15   & $4644-0.5$                 & $4645-0.4i$     \\
 3.20   & $4642-0.8i$    & $4628-1.8i$   \\
 3.25   & $4640-1.2i$    & $4600-4.5i$ \\
 3.35   & $4632-2.6i$    & $--$     \\ 
 3.45   & $4618-4.7i$  & $--$      \\
 3.55   & $--$  & $--$     \\
\toprule[1.5pt]
\end{tabular}
\end{center}

\end{table}

The results suggest that there are only two states produced from the  interaction ${D}^*_s\bar{D}_{s1}-{D}_s\bar{D}_{s1}$ with cutoff in a range from 3.05 to 3.55 GeV with binding energy smaller than 50 MeV.  The both states are near ${D}^*_s\bar{D}_{s1}$ threshold with spin parity  $J^P=1^{-}$ but with different  charge parities, positive $C=1$ and negative $C=-1$. It is found that these two states appear at a cutoff $\Lambda$ about 3.1 GeV.  It  suggests that the effect of  cross diagram is relatively small especially near the threshold  as shown in Table~\ref{flavor factor} because the $C$ only involves in the contribution from cross diagram. With the increase of the cutoff, the difference between two states becomes larger. The binding energy of the state with $C=-1$ is relatively stable, and reaches 50 MeV at cutoff about 3.5 GeV.  For the $C=1$ state, the binding energy increases rapidly to 50 MeV at a cutoff of 3.25 GeV. If we adopt a cutoff of $3.4$ GeV, which leads to a bound state with $J^{PC}=1^{--}$  at $z=4626-3.4i$ MeV as shown in Fig.~\ref{pole} , the binding energy of the state with $1^{-+}$ will be large.  However, considering the uncertainties in theory and experiment,    if the $1^{--}$ state can be related to the $Y(4626)$, the $1^{-+}$ state is still promising to be observed in experiment.

Now, we loose the condition, and find the bound states in a larger range of $0.5<\Lambda<5$ GeV  for reference. Here, we still only consider the states with binding energy smaller than 50 MeV. With such condition, more states can be found,  and  in Table~\ref{Tab: bound state2} we list the results in a range of cutoff from 2.3 to 3.1 GeV because no bound states with binding energy smaller than 50 MeV can be found at cutoffs below 2.3 GeV and larger than 3.1 GeV. It is found that the $2^{-}$ and $0^{-}$ states can  be also formed in S-wave from the ${D}^*_s\bar{D}_{s1}$ system. The bound states  with $2^-$  appear at cutoff $\Lambda=2.9$ and 2.8 GeV for negative and positive charge parities, respectively. Compared with the case with spin parity $1^{--}$, the binding energies increase very rapidly to larger than 50 MeV, which is beyond the scope of a molecular state. For the $0^{-}$ state,  the bound states are found for both positive and negative charge parities at smaller cutoffs. Because the quantum number $0^-$ is forbidden for the ${D}_s\bar{D}_{s1}$ system, no width is produced for these states in our two-channel calculation. Below the ${D}_s\bar{D}_{s1}$ threshold, $1^-$ states are also produced with both charge parities.  We would like to remind that  existence of these six states needs much smaller cutoffs compared with two ${D}^*_s\bar{D}_{s1}(1^-)$ states.

\renewcommand\tabcolsep{0.19cm}
\renewcommand{\arraystretch}{1.51}
\begin{table}[h!]
\begin{center}
\caption{More bound states from the interaction ${D}^*_s\bar{D}_{s1}-{D}_s\bar{D}_{s1}$  at different cutoffs $\Lambda$  in a range from 0.5 to 5 GeV. Here, ``$--$" means that no bound state is found or the bound state has a binding energy larger than 50 MeV.
The cutoff $\Lambda$ and  position of the pole $z$ are in units of GeV and MeV, respectively. \label{Tab: bound state2}
\label{diagrams}}
	\begin{tabular}{c|cccc|cc}\bottomrule[1.5pt]
&\multicolumn{4}{c}{${D}^*_s\bar{D}_{s1}$}&\multicolumn{2}{|c}{${D}_s\bar{D}_{s1}$}\\\hline
$\Lambda$ & $2^{--}$& $2^{-+}$& $0^{--}$& $0^{-+}$& $1^{--}$& $1^{-+}$
\\\hline
2.30  &$--$             & $--$             &$--$        & $--$    &$--$      & $--$     \\
2.40  &$--$             & $--$             &$--$        & $4646$  &$--$      & $--$   \\
2.45  &$--$             & $--$             &$--$        & $4644$  &$--$      & $--$      \\
2.50  &$--$             & $--$             &$--$        & $4639$  &$--$      & $--$   \\
2.55  &$--$             & $--$             &$--$        & $4632$  &$4496$      & $--$  \\
2.60  &$--$             & $--$             &$4647$    & $4621$   &$4486$      & $--$   \\
2.65  &$--$             & $--$             &$4643$     & $4604$  &$4478$      & $--$   \\
2.70  &$--$             & $--$             &$4633$     & $--$     &$4471$      & $--$  \\
2.75  &$--$             & $--$             &$4614$     & $--$   &$4466$      & $--$   \\
2.80  &$--$             & $4647-2.0i$  &$--$        & $--$   &$4461$      & $4503$  \\
2.85  &$--$             & $4643-6.6i$  &$--$        & $--$      &$4453$    & $4487$  \\
2.90  &$4646-1.0i$   & $4598-5.5i$  &$--$        & $--$    &$--$      & $4476$    \\
2.95  &$4643-1.6i$   & $--$            &$--$        & $--$    &$--$      & $4461$   \\
3.00  &$4634-1.6i$   & $--$            &$--$        & $--$     &$--$      & $--$    \\
3.05  &$4602-0.4i$  & $--$            &$--$        & $--$     &$--$      & $--$  \\
3.10  &$--$             & $--$            &$--$        & $--$    &$--$      & $--$   \\
\toprule[1.5pt]
\end{tabular}
\end{center}

\end{table}

\section{Summary and discussion}

Inspired by the newly observed $Y(4626)$, we study the possible ${D}^*_s\bar{D}_{s1}$ molecular state in a qBSE approach  within the one-boson-exchange model.  A two-channel calculation of the ${D}^*_s\bar{D}_{s1}-{D}_s\bar{D}_{s1}$ interaction is performed to search for the pole produced from the interaction. 
A state with quantum numbers $J^{PC}=1^{--}$ can be produced at about 4626 MeV near the ${D}^*_s\bar{D}_{s1}$  threshold, it can be related to the $Y(4626)$ observed at Belle recently. Such molecular state couples with the  ${D}_s\bar{D}_{s1}$ channel through exchange $\phi$ and $\eta$ mesons. Our result shows that a peak around 4626 MeV is produced in the ${D}_s\bar{D}_{s1}$ invariant mass spectrum, which corresponds to the pole from the interaction ${D}^*_s\bar{D}_{s1}-{D}_s\bar{D}_{s1}$. Hence, it is consistent with the observation of the $Y(4626)$ at the  ${D}_s\bar{D}_{s1}$  channel. However, the width obtained theoretically  is smaller than the experimental one with reasonable parameters, which suggests that other channels may provide important contribution to the total width of the  $Y(4626)$.

Besides the state corresponding to the $Y(4626)$, we also give  prediction  of other possible molecular states from the ${D}^*_s\bar{D}_{s1}-{D}_s\bar{D}_{s1}$ interaction.  Based on our result, under the assumption that the $Y(4626)$ is a molecular state from the interaction ${D}^*_s\bar{D}_{s1}(2536)-{D}_s\bar{D}_{s1}(2536)$, the most promising state is the $1^{-+}$ state, which is different from the $1^{--}$ state corresponding to the $Y(4626)$ only in the charge parity, and produced at almost the same cutoff as $1^{--}$ state.  The $0^-$ and $2^-$ states are also found near the ${D}^*_s\bar{D}_{s1}$ thresholds.  The values to produce these states are smaller than the one to reproduce the $Y(4626)$, which suggests that these four states are not so reliable as the $1^{-+}$ state. More theoretical and experimental works are required to clarify their existence. These two states can not couple with ${D}_s\bar{D}_{s1}$ channel in S wave, and should not have obvious effect on the ${D}_s\bar{D}_{s1}$ invariant mass spectrum, where the $Y(4626)$ was observed. For the lower  ${D}_s\bar{D}_{s1}$ threshold, there are also $1^-$ states produced, which can be seen as the partner of the $Y(4260)$. For these two  states from the interaction ${D}_s\bar{D}_{s1}$, the cutoff can be different from 3.4 GeV adopted for the ${D}^*_s\bar{D}_{s1}$ interaction. Hence, its existence may be more reliable than the $0^-$ and $2^-$ states from the ${D}^*_s\bar{D}_{s1}$  interaction, which need be checked by future experimental observation.  It is interesting to see that three of the predicted states, the $1^{-+}$ and $0^{--}$ states from the  ${D}^*_s\bar{D}_{s1}$ interaction and $1^{-+}$ state from the  ${D}^*_s\bar{D}_{s1}$ interaction, carry exotic quantum numbers. Based on these results, we strongly suggest to search for the $1^{-+}$ state near the  ${D}^*_s\bar{D}_{s1}$ threshold in future experiment.

\vskip 10pt

\noindent {\bf Acknowledgement} This project is supported by the National Natural Science
Foundation of China (Grants No. 11675228, and No. 11775050), and the Fundamental Research Funds for the Central Universities.

%


\end{document}